\documentclass[12pt,showpacs,aps]{revtex4}

\usepackage{graphicx}
\begin{document}
\title{Influence of qubit displacements on quantum logic operations\\ 
in a silicon-based quantum computer with constant interaction}
\author{D. I. Kamenev$^1$, G. P. Berman$^1$,  and V. I. Tsifrinovich$^2$}
\affiliation{$^1$Theoretical Division, T-13,
Los Alamos National Laboratory, Los Alamos, New
Mexico 87545}
\affiliation{$^2$ Department of Physics, Polytechnic University,
Brooklyn, New York 11201}

\begin{abstract}
The errors caused by qubit displacements from their prescribed locations
in an ensemble of spin chains are estimated analytically and
calculated numerically for a quantum computer based on phosphorus
donors in silicon. We show that
it is possible to polarize (initialize) the nuclear spins even with
displaced qubits by using Controlled NOT gates between
the electron and nuclear spins of the same phosphorus atom.
However, a Controlled NOT gate between the
displaced electron spins is implemented with large error because
of the exponential dependence of
exchange interaction constant on the distance between the qubits.
If quantum
computation is implemented on an ensemble of many spin chains, the
errors can be small if the number of chains with displaced qubits is small.
\end{abstract}
\maketitle

\section{Introduction}
A promising candidate for solid-state quantum computation
is the phosphorus-doped silicon. This semiconductor material is the
backbone of microelectronic technology. In order for this architecture
to be useful for quantum information processing, the positions of the
phosphorus dopants in silicon must be controlled by utilizing, for
example, a scanning tunneling microscope
(STM)~\cite{1994,1996,surfSci,electronics,clark}. The STM can be
used to create many identical arrays of phosphorus atoms on
the surface of silicon. A large number of such arrays is required
for the detection of the qubit states. The long decoherence time of the nuclear
spins of the phosphorus donors in silicon makes this quantum computer
attractive for quantum information processing.

Kane~\cite{Kane} proposed using nanoscale electronic gates to
control the qubits. This technique has not yet been 
experimentally realized, so in
this paper we consider a different architecture. In our approach,
the exchange interaction between qubits is constant, and selective
interactions are realized through the use of a magnetic field
gradient and both microwave and radio-frequency pulses. Measurement
can be implemented using optical techniques similar to those used in
Refs.~\cite{optical1,optical2,optical3}.

Each phosphorus atom has a nuclear spin 1/2 and an electron spin 1/2.
There is a hyperfine interaction between both spins. The information
is stored in the states of the nuclear spins. The interaction between
the nuclear spins of neighboring phosphorus atoms is mediated by
electron spins coupled to each other by the exchange interaction.  The
advantage of an architecture based on controlled phosphorus
impurities in silicon are the potential scalability and the possibility of
using advanced silicon-based semiconductor technology in the 
quantum computer design.

The typical size of an STM tip is larger than the lattice constant
of silicon. Consequently, the phosphorus atoms in the lattice can be
shifted from their prescribed locations by 1-4 lattice sites.
Errors are generated because the Larmor frequencies of the displaced
spins  and the exchange interaction constant are modified. In this
paper, this error is estimated analytically and calculated
numerically for the protocol required to polarize the nuclear spins
and for Controlled NOT gate between the electron spins of neighboring
phosphorus atoms. If realized experimentally, each qubit chain in
the first quantum silicon-based quantum computer 
will possibly contain the minimum
number (two) of coupled qubits required to demonstrate basic
principles of quantum computation in semiconductors. Hence, we
specialize in this paper to the case of two phosphorus atoms in a
chain. The unwanted displacements of the qubits seem to be
inevitable in the quantum computer architecture. Therefore, our results are
important for evaluating the possibility of constructing a working
silicon-based quantum computer.

\section{Hamiltonian}
\begin{figure}
%\vspace{-7mm}
\includegraphics[width=10cm,height=8cm,clip=]{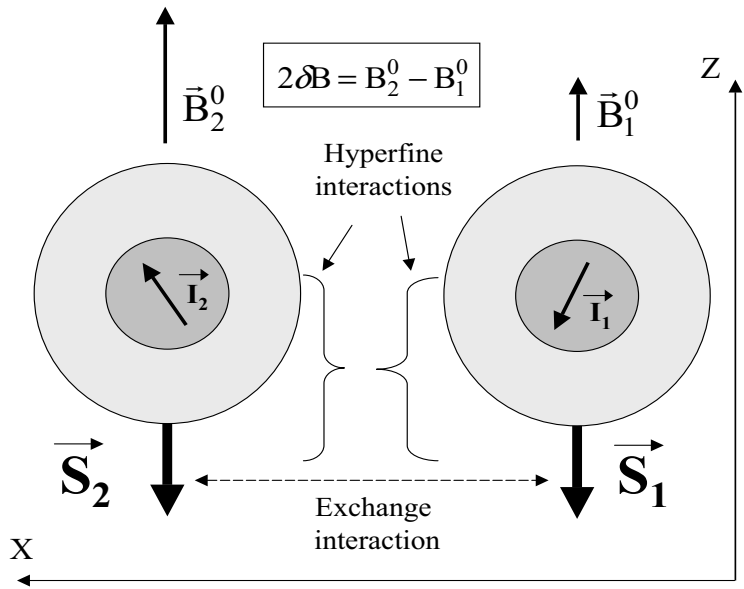}
\vspace{-4mm}
\caption{A schematic illustration of two phosphorus atoms placed
in a permanent magnetic field. The electron spins $\vec S_1$ and
$\vec S_2$ (large arrows) of the neighboring atoms
interact with each other via the exchange interaction
and the nuclear spins $\vec I_1$ and
$\vec I_2$ (small arrows) interact with the electron spins
through the hyperfine interactions.}
\label{reffig:fig1}
\end{figure}

A schematic illustration of the system under consideration is given
by Fig.~\ref{reffig:fig1}. If the qubits in each chain are placed at a
separation of $\sim$ 35 nm from each other, the nuclear-nuclear 
($\sim 3.5\times 10^{-4}$ Hz),
nuclear-electron ($\sim 0.65$ Hz), 
and electron-electron ($\sim 1.2$ kHz) dipole-dipole interactions
are small compared to the electron-electron exchange interaction 
($\sim 2$ MHz), so 
that one can neglect the dipole-dipole interactions.
Since the relaxation time for the electron spins at
temperatures of 1-7 K is relatively short (0.6-60 ms at 7
K~\cite{T_c}), the quantum information must be stored in the states
of the nuclear spins. Because the nuclear spins do not interact
directly, electron spins can be used to mediate the nuclear-nuclear
interactions. In this setup, the electron spins must be coherent
only during the relatively short time of implementation of the quantum
logic gate, such as a Controlled-NOT gate on a particular pair of qubits.
The coupled electron spins of the neighboring phosphorus atoms 
serve as the static ``quantum bus'' which transfers the interaction between 
nearest-neighbor and distant nuclear spins in a chain with more than two 
phosphorus atoms. In a similar technique ``flying'' qubits represented 
by photons has been proposed to mediate the qubit-qubit 
interaction between 
trapped ions~\cite{obus1,obus2,obus3,obus4,obus5,obus6,obus7}. 

The Hamiltonian is
\begin{equation}
\label{H}
\hat H=\hat H^0+\hat V(t),
\end{equation}
where $\hat H^0$ is the unperturbed Hamiltonian of the
system shown in Fig.~\ref{reffig:fig1} and $\hat V(t)$ is the
time-dependent field describing rectangular radio-frequency
pulses used to implement quantum logic gates.
The Hamiltonian $\hat H^0$ is
\begin{equation}
\label{H0} \hat H^0=\gamma_e B_1^z\hat S_1^z+\gamma_e B_2^z\hat
S_2^z- \gamma_n B_1^z\hat I_1^z-\gamma_n B_2^z\hat I_2^z+
A\left(\hat{\vec S}_1\hat{\vec I}_1+\hat{\vec S}_2\hat{\vec
I}_2\right)+ J\hat{\vec S}_1\hat{\vec S_2},
\end{equation}
where $\hat{\vec I}_k$ and $\hat{\vec S}_k$ are the spin operators
of, respectively, the $k$th nuclear and electron spins, $k=1,2$; $\hat
I_k^z$ and $\hat S_k^z$ are the projections of these operators on
the $z$ axis; $\gamma_e$ and $\gamma_n$ are, respectively, the
magnitudes of the electron and nuclear gyromagnetic ratios; $B_k^0$
is the permanent magnetic field at the location of the $k$th spin;
$A$ and $J$ are, respectively, the hyperfine and exchange
interaction constants.

We utilized perturbation theory to calculate the
eigenvalues $E_i$, $i=0,1,\dots,15$ of the Hamiltonian $\hat H^0$ which are
given in Appendix~\cite{xxx}. In spite of the fact that our system
allows exact analytical calculation of $E_i$, we prefer to use a perturbation
approach because this technique allows us to separate the most important
contributions $E_i^{(0)}$ to $E_i$ from the less important
and to use relatively simple expressions for the eigenvalues to
analyze the dynamical properties of this system. The small corrections
$E_i^{(2)}$ are useful for calculating the parameters of the
pulses in numerical simulations. The eigenvalues $E_i$, $i=1,2\dots,14$
are calculated with an accuracy of $\xi^2(A/2)/(2\pi)\approx 23$ Hz,
where
$$
\xi={A\over 2\gamma_e b}\approx 6\times 10^{-4}.
$$
Here we assume $b=(B_1^0+B_2^0)/2=3.3$ T,
so that  $\gamma_eb/(2\pi)=92.5$ GHz, $\gamma_e/(2\pi)=28.025$ GHz/T,
$A/(2\pi)=117.53$ MHz~\cite{feher}.

The basis states are
\begin{equation}
\label{basis}
|n_2e_2e_1n_1\rangle,
\end{equation}
where the electron spin $e_i$ and nuclear spin $n_i$ of the $i$th
phosphorus atom, $i=1,2$, can assume the values 0 and 1; the state
$|0\rangle$ corresponds to the orientation of the spin along the
direction of the permanent external magnetic field and the state
$|1\rangle$ corresponds to the opposite direction. The state
$|0000\rangle$ (which is not the ground state) has energy $E_0$,
the state $|0001\rangle$ has energy $E_1$, etc. In the spin chain
(\ref{basis}) there are interactions only between the neighboring
spins, so this kind of spin ordering is convenient for the analysis
of conditional quantum logic gates.

The basis states (\ref{basis}) are not the eigenstates of the
Hamiltonian $\hat H_0$ because of the (off-diagonal) terms
$AI^x_kS^x_k$, $AI^y_kS^y_k$, $k=1,2$, and $JS^x_1S^x_2$,
$JS^y_1S^y_2$. However, the eigenfunctions approximately coincide
with the basis states~(\ref{basis}) if the conditions
\begin{equation}
\label{epsilon}
\epsilon={J\over 2\gamma_e(B_2^z-B_1^z)}\ll 1,~~~
\epsilon^\prime={J\over |2\gamma_e(B_2^z-B_1^z)-A|}\ll 1,~~~
\xi\ll 1
\end{equation}
are satisfied~\cite{xxx}. The unwanted changes (errors)
in the wave function due to the influence of the off-diagonal terms
are of the order of $\epsilon$ or $\epsilon^\prime$ or $\xi$.

\section{Implementation of quantum logic gates}
The time-dependent magnetic field has the following components:
\begin{equation}
\label{B1t}
\vec B^1(t)=B^1(\cos(\nu t+\varphi),-\sin(\nu t+\varphi),0),
\end{equation}
where $B^1$, $\nu$, and $\varphi$ are, respectively, the amplitude,
frequency and phase of the pulse, and $t$ is time. The frequency $\nu$
can assume both positive and negative values~\cite{xxx}.
The time-dependent term in the Hamiltonian has the form
$$
\hat V(t)=\left[{\Omega_e^0\over 2}\left(\hat S_1^-+\hat S_2^-\right)-
{\Omega_n^0\over 2}\left(\hat I_1^-+\hat I_2^-\right)\right]e^{-i(\nu
t+\varphi)}+
$$
\begin{equation}
\label{Vt}
\left[{\Omega_e^0\over 2}\left(\hat S_1^++\hat S_2^+\right)-
{\Omega_n^0\over 2}\left(\hat I_1^++\hat I_2^+\right)\right]e^{i(\nu
t+\varphi)},
\end{equation}
where $\Omega_e^0=\gamma_e B^1$ and $\Omega_n^0=\gamma_n B^1$.

We will consider the following three basic kinds of
quantum gates: a Controlled NOT gate with the control nuclear spin and
the target electron spin of the same phosphorus atom, a Controlled NOT
gate with the control electron spin and the target nuclear spin, and
a Controlled NOT gate between the neighboring electron spins. We now
briefly describe the procedure for calculating the optimal parameters of the
pulses required to implement these gates. (For a detailed analysis
see Ref.~\cite{xxx}.) Assume that the direction of the $k$th spin in
the state $|p\rangle$ is along the direction of the permanent
magnetic field $\vec B^0_k$, $k=1,2$ 
(i.e. $|p\rangle=|\cdots 0_k\cdots\rangle$) 
and the state $|q\rangle=|\cdots 1_k\cdots\rangle$ is related to the state
$|p\rangle$ by a flip of the $k$th spin.
In order to flip the $k$th
spin in the state $|p\rangle$ or in the state $|q\rangle$, the
frequency $\nu$ and the time-duration $\tau$ of the rectangular
pulse must satisfy the conditions
\begin{equation}
\label{nu_tau}
\nu=E_q-E_p,~~~\tau={\Omega\over\pi},
\end{equation}
where $\Omega$ is the Rabi frequency of the pulse.

\begin{figure}
\includegraphics[width=9cm,height=5cm,clip=]{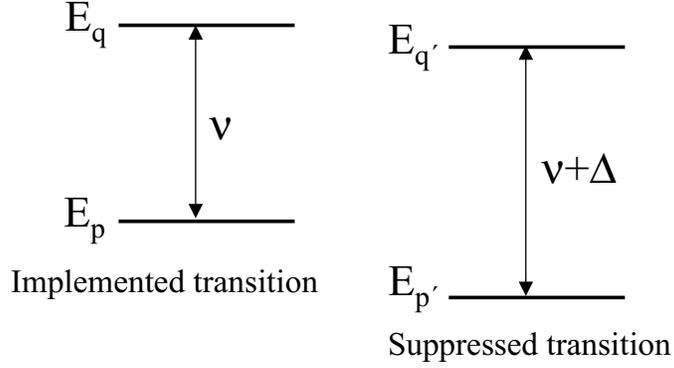}
\vspace{-4mm}
\caption{A schematic illustration of the energy levels ivolved in 
implementation of a Controlled-NOT gate. The frequency
$\nu$ is the frequency of the pulse.}
\label{reffig:transitions}
\end{figure}

Assume that we wish to suppress the transition
$|p^\prime\rangle\leftrightarrow|q^\prime\rangle$, where
the $k$th spin in the state $|p^\prime\rangle$
is along the direction of the permanent magnetic field $\vec B^0_k$
and the state $|q^\prime\rangle$ is related to the state $|p^\prime\rangle$
by a flip of the $k$th spin; $|p\rangle\ne |p^\prime\rangle$ and
$|q\rangle\ne |q^\prime\rangle$. A schematic illustration of the involved
energy levels and transitions is shown in Fig.~\ref{reffig:transitions}.
The Rabi frequency of the pulse
must satisfy the $2\pi K$-condition~\cite{book,book1}
\begin{equation}
\label{Omega_e}
\Omega={|\Delta|\over\sqrt{4K^2-1}},
\end{equation}
where $K=1,2,\dots$ is an integer number,
$\Omega=\Omega_e$ for an electron spin and
$\Omega=\Omega_n$ for a nuclear spin, and
\begin{equation}
\label{Delta_b}
\Delta=E_{q^\prime}-E_{p^\prime}-\nu.
\end{equation}
The probability of transition generated by the pulse with the Rabi frequency
$\Omega$ and detuning $\Delta$ is~\cite{book,book1}
\begin{equation}
\label{R}
R={\Omega^2\over \lambda^2}
\sin^2\left({\pi\lambda\over 2\Omega}\right),
\end{equation}
where $\lambda=\sqrt{\Omega^2+\Delta^2}$. We have $R=1$ for 
$\Delta=0$ and $R=0$ if $\Omega$ satisfies Eq.~(\ref{Omega_e}).
These two conditions allow one to implement the Controlled 
NOT gate as indicated in Fig.~\ref{reffig:transitions}.

\section{$J$ as a function of the distance between the qubits}
A simple approximation of the exchange interaction adequate for
our purposes is given by the Herring-Flicker
formula~\cite{Kane,HerringFlicker}
\begin{equation}
\label{J}
J=1.642{e^2\over \kappa a_{\rm B}}\left({a\over a_{\rm B}}\right)^{5/2}
\exp\left(-{2a\over a_{\rm B}}\right),
\end{equation}
where $e$ is the electron charge, $\kappa=11.9$ is the
dielectric constant of silicon,
and $a$ is the distance between the qubits. The effective
Bohr radius is
\begin{equation}
\label{aB} a_{\rm B}=\kappa{M\over M^*}a_{\rm B}^0\approx
33.14~\text{\AA},
\end{equation}
where $a_{\rm B}^0\approx 0.5292$~\AA~is the Bohr radius, $M$ is the
electron mass, $M^*=0.19M$ is the effective electron mass.
In Eq.~(\ref{J}), $a=Na_0$ where $N-1$ is the number
of interstitial silicon
atoms between the two phosphorus atoms serving as qubits
and $a_0=7.68$~\AA. We assume that the qubits are placed on the (100)
surface of silicon and the direction of the permanent magnetic field is
perpendicular to this surface.
In Table I we show the values of $J/(2\pi)$
for different $N$ calculated using Eq.~(\ref{J}).
The controlled separation between the qubits can be varied by changing $N$.
In this paper we assume that the predefined spacing
(desired spacing without unwanted displacements) is
$N_0=47$, so that the predefined
exchange interaction constant is $J_0/(2\pi)=1.97$ MHz.

\begin{table}[th]
\begin{tabular}{|c|c|c|c|c|c|c|c|c|c|c|c|c|}\hline
\label{table1}
$N$&40&41&42&43&44&45&46&47&48&49&50&51\\
$a$ (nm)& 30.72 & 31.49 & 32.26 & 33.02 & 33.79 & 34.56 & 35.33 &
36.10 & 36.86 & 37.63 & 38.40 & 39.17\\
$J/(2\pi)$ (MHz)&~33.75~&~22.58~&~15.09~&~10.07~&
~6.71~&~4.465~&~2.97~&~1.97~&~1.306~&~0.865~&~0.573~&~0.37855
\\  \hline
\end{tabular}
\caption{The distance $a$ between the
neighboring phosphorus atoms and the
exchange interaction constant
$J/(2\pi)$ for different $N$, where
$N-1$ is the number of interstitial silicon atoms.}
\end{table}

\section{Controlled NOT gates between the electron and nuclear spins}
We now derive the pulse parameters of four pulses required to
initialize the nuclear spins. Each pulse implements
one logical gate. The gates are~\cite{xxx}
\begin{equation}
\label{Initialization}
{\rm (a)}~{\rm CN_{n1,e1}},~~~{\rm (b)}~{\rm CN_{e1,n1}},~~~
{\rm (c)}~{\rm CN_{n2,e2}},~~~{\rm (d)}~{\rm CN_{e2,n2}},
\end{equation}
where CN$_{i,j}$ denotes a Controlled NOT gate, $i$ is the number 
of the control
qubit and $j$ is number of the target qubit. The protocol~(\ref{Initialization})
includes also a delay time (of the order of 0.1 ms)
between pulses (b) and (c) and after the pulse (d).
During this time the electron spins relax to their ground states.
We assume that the relaxing electron spins do not influence
the corresponding nuclear spins~\cite{xxx}.

In order to calculate the energies and resonant frequencies
we will use the following parameters:
$$
{J/2\over 2\pi}=0.985~{\rm~MHz},~~~
{A/2\over 2\pi}=58.765~{\rm~MHz},~~~
{\gamma_e b\over 2\pi}=92.48~{\rm~GHz},~~~
{\gamma_n b\over 2\pi}=56.93~{\rm~MHz},
$$
\begin{equation}
\label{parameters}
{\gamma_e \delta B\over 2\pi}=65.76~{\rm~MHz},~~~
{\gamma_n \delta B\over 2\pi}=40.48~{\rm~kHz},
\end{equation}
where $\delta B=(B_2-B_1)/2$.
Here we assume $b=3.3$ T, $\gamma_n/(2\pi)=17.25144$ MHz/T,
the magnetic field gradient is $1.3\times10^5$ T/m,
and from Table I we use $a=36.1$ nm.

We now calculate the parameters of the pulse implementing gate (a).
The resonant transition is
$|1101\rangle\leftrightarrow |1111\rangle$. The resonant frequency
for this transition is
\begin{equation}
\label{nul}
{\nu_1\over 2\pi}\approx{E_{15}-E_{13}^{(0)}\over 2\pi}={1\over 2\pi}\left[
-\gamma_e b+{A\over 2}+{J\over 2}+\sqrt{(\gamma_e\delta B)^2+J^2/4}\right]
=-92.35~{\rm GHz}.
\end{equation}
The near-resonant transition to be suppressed is
$|1100\rangle\leftrightarrow |1110\rangle$. The detuning for this
transition is
$$
{\Delta_1\over 2\pi}\approx{E_{14}^{(0)}-E_{12}^{(0)}-\nu_1\over 2\pi}=
$$
\begin{equation}
\label{Delta1}
{1\over 2\pi}\left[\sqrt{(\gamma_e\delta B-A/2)^2+J^2/4}-
\sqrt{(\gamma_e\delta B)^2+J^2/4}-{A\over 2}\right]=-117.47~{\rm MHz}.
\end{equation}
In order to suppress the near-resonant transition with the detuning
$\Delta_i$ the Rabi frequency of the pulse must satisfy the $2\pi K$
condition
\begin{equation}
\label{OmegaI}
{\Omega_i\over 2\pi}=
{|\Delta_i|\over 2\pi\sqrt{4K^2-1}},
\end{equation}
where $K$ is an integer number and $i$ is the pulse number.
For the first pulse with $\Delta_1$
given by Eq.~(\ref{Delta1}) and for $K=K_e=1$, we have
\begin{equation}
\label{Omega1}
{\Omega_1\over 2\pi}={|\Delta_1|\over 2\pi\sqrt{3}}=67.82~{\rm MHz}.
\end{equation}

\begin{figure}
%\vspace{-7mm}
\includegraphics[width=12cm,height=8cm]{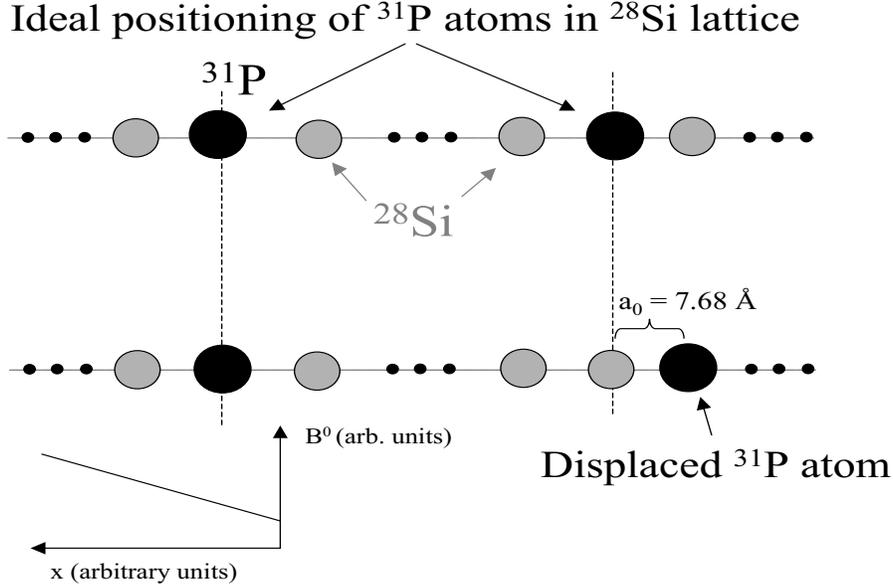}
\vspace{-4mm}
\caption{A schematic illustration of an ideal qubit chain (upper part)
and a qubit chain with a displaced qubit (lower part).}
\label{reffig:fig2}
\end{figure}

There are three sources of errors in our model: (i) nonresonant
excitations, (ii) off-diagonal components of the exchange and hyperfine interactions, and (iii) qubit displacements. The error in the probability amplitude
due to the nonresonant transitions is proportional to the ratio~\cite{PR02}
\begin{equation}
\label{mu}
\mu={\Omega\over 2|\delta\omega|},
\end{equation}
where $\delta\omega$ is the difference
between the frequency corresponding to the nonresonant transition and
the frequency of the pulse.

The order of magnitude of the error generated by a single pulse
acting on an electron spin
can be estimated by adding probability errors generated due to all three
mechanisms (i)-(iii),
\begin{equation}
\label{P}
P=1-R+\epsilon^2+\mu^2,
\end{equation}
where $\epsilon$, $R$, and $\mu$ are given by Eqs.~(\ref{epsilon}), (\ref{R}),
and (\ref{mu}).

We now analytically estimate the error generated by the first pulse
due to the displacement of the first phosphorus atom using
Eq.~(\ref{R}). This displacement introduces an unwanted change in
the Larmor frequency of the first qubit and the correction $dJ$ to the
exchange interaction constant $J$. Let qubit 1 be displaced in the
negative $x$ direction by one lattice site as shown in
Fig.~\ref{reffig:fig2}. From Table I we find
\begin{equation}
\label{dJ}
{dJ\over 2\pi}=1.306~{\rm MHz} -1.97~{\rm MHz}=-0.664~{\rm MHz}.
\end{equation}
The permanent magnetic field
at the location of the first qubits is $B_1^0-dB$, where
\begin{equation}
\label{dB}
dB={B_2^0-B_1^0\over N}\approx  9.985\times10^{-5}~{\rm T}.
\end{equation}
From Eq.~(\ref{nul}) we have 
$$
\nu_1\approx -\gamma_eB_1^0+{A\over 2}+{J\over 2},
$$
so that the unwanted detuning is 
\begin{equation}
\label{Delta_primeI}
{\Delta_1^\prime(m=-1)\over 2\pi}\approx {1\over 2\pi}
\left(\gamma_edB +{\delta J\over 2}\right)
=2.466~{\rm MHz}, 
\end{equation}
where $\gamma_edB/(2\pi)=2.798$ MHz and
we used Eqs.~(\ref{dJ}) and (\ref{dB}).
If we put $\Delta^\prime_1$ instead of $\Delta$ and $\Omega_1$
instead of $\Omega$ in Eq.~(\ref{R}) we find the probability error $1-R$
due to the unwanted qubit displacement. 

For displacement in the opposite direction with $m=1$
\begin{equation}
\label{Delta_primeIl}
{\Delta_1^\prime(m=1)\over 2\pi}\approx {1\over 2\pi}\left(-\gamma_edB +
{\delta J^\prime\over 2}\right)=-1.8~{\rm MHz},
\end{equation}
where $\delta J^\prime/(2\pi)=1$ MHz (see Table I). Since 
$|\Delta_1^\prime(m=1)|<|\Delta_1^\prime(m=-1)|$, 
the error $P_1$ for the displacement in the negative direction along 
the $x$ axis is larger than the error for displacement in the positive 
direction because $1-R$ in Eq.~(\ref{P}) increases with $|\Delta_1^\prime|$
increasing. 

\begin{figure}
%\vspace{-7mm}
\includegraphics[width=10cm,height=8cm]{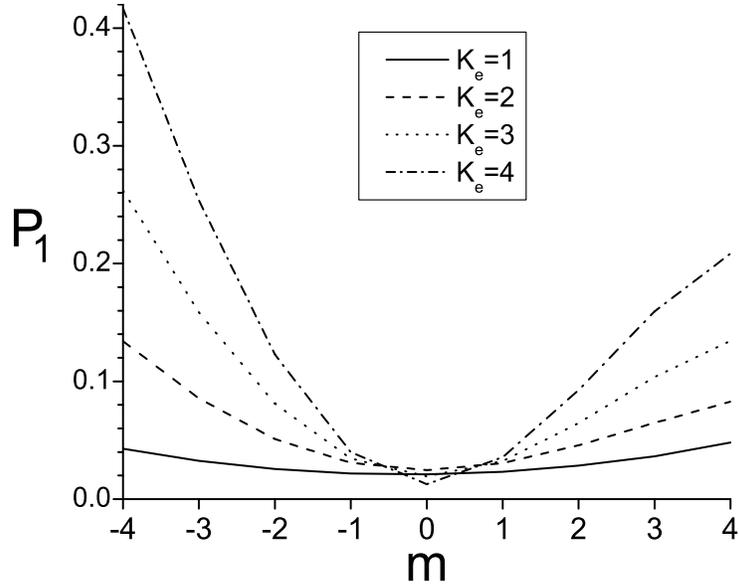}
\vspace{-4mm}
\caption{Probability error $P_1$ generated by gate (a) in 
Eq.~(\ref{Initialization}) as a function of displacement $m$ of the first
phosphorus atom for four values of $K_e$.}
\label{reffig:P1}
\end{figure}

In Fig.~\ref{reffig:P1} we plot the error $P_1$ for the
transition $|1101\rangle\rightarrow|1111\rangle$ generated by the
first pulse for the chain with displaced qubit 1 
as a function of the qubit displacement $m$ along
the $x$ axis. This plot is asymmetric 
with respect to $m=0$ which follows from the
analysis presented in the previous paragraph.
The data are obtained by utilizing the eigenfunctions
of full Hamiltonian in the rotating frame. (For a description of the numerical
solution see Ref.~\cite{xxx}.) The parameters of the pulses in our
numerical simulations are calculated analytically using the
eigenvalues from the Appendix. In the figure, $m$ is the number
of lattice sites for a displaced qubit, so that the displacement is
$ma_0$, $m=-4,-3,\dots,4$. Our simulations (not presented here) show
that this and other pulses generate the maximum error in situation when
the ``active'' qubit to be flipped is displaced and the error is smaller 
when the neighboring qubit is displaced. 

Since $|\Delta^\prime_1|\ll \Omega_1$
[see Eqs.~(\ref{Omega1}), (\ref{Delta_primeI}), and (\ref{Delta_primeIl})]
the error $P_1$ is relatively small small. (Here and below we assume that 
the error $P$ is small if $P\ll 1$.)
The smallness of the error generated by the first pulse is due to 
(a) the neighboring phosphorus atoms being located relatively far from 
each other; 
(b) the electron Rabi frequency being relatively large, and (c) the exchange
interaction constant $A$, responsible for the Controlled NOT gate, 
not being modified by the displacements of the phosphorus atoms.  

\begin{figure}
%\vspace{-7mm}
\includegraphics[width=10cm,height=8cm]{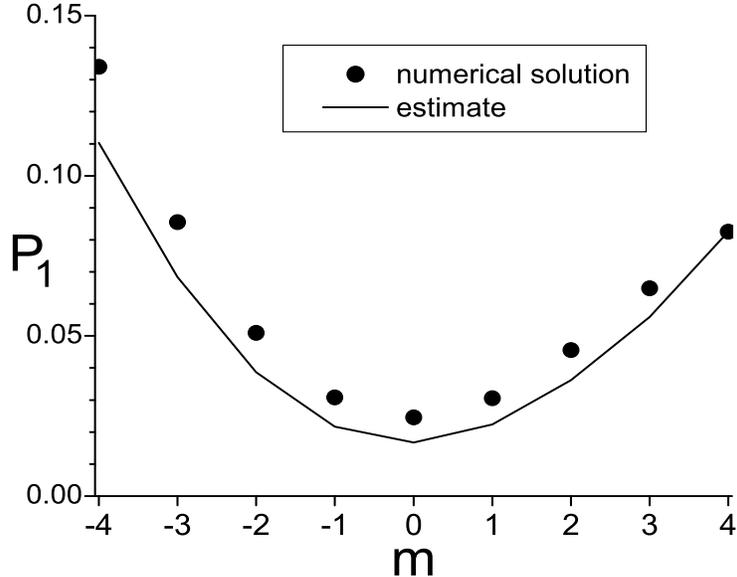}
\vspace{-4mm}
\caption{Probability error $P_1$ generated by by gate (a) in 
Eq.~(\ref{Initialization}) as a function of displacement $m$ of the first
phosphorus atom along the $x$ axis for $K_e=2$.
The estimate is calculated using Eq.~(\ref{P}).}
\label{reffig:k4}
\end{figure}

From Fig.~\ref{reffig:P1}, one can see that the error increases
as $|m|$ increases. As follows from this figure, the
probability error becomes more sensitive to the qubit displacement
when $K_e$ becomes relatively large and the Rabi frequency becomes
relatively small. In Fig.~\ref{reffig:k4} the probability error
estimate (\ref{P}) is compared with the results of the numerical solution.
One can see that our estimate is satisfactory and supports the
approach taken. The asymmetry of the plot in Fig.~\ref{reffig:k4}
with respect to $m=0$ is of the same origin as that in Fig.~\ref{reffig:P1}
discussed before.

We now estimate the error associated with a flip of the first nuclear spin
in the situation when the first phosphorus atom is shifted from its ideal
position to $m=-1$.
We first calculate the pulse parameters. The resonant transition is
$|1111\rangle\leftrightarrow |1110\rangle$. The frequency of this transition is
\begin{equation}
\label{nu2}
{\nu_2\over 2\pi}\approx{E_{15}-E_{14}^{(0)}\over 2\pi}={1\over 2\pi}\left(
\gamma_n b-\gamma_n \delta B + {A\over 2}\right)
=115.655~{\rm MHz}.
\end{equation}
The near-resonant transition to be suppressed is
$|1100\rangle\leftrightarrow |1101\rangle$. One can show that
the detuning for this transition is $\Delta_2=\Delta_1$, where
$\Delta_1$ is given by Eq.~(\ref{Delta1}).
The $2\pi K$ condition for the second pulse reads
\begin{equation}
\label{Omega2}
{\Omega_2\over 2\pi}=
{|\Delta_2|\over 2\pi\sqrt{4K_n^2-1}},
\end{equation}
where $K_n$ is an integer. In order to suppress the flip of
nuclear spin 2, the value of $\Omega_2$ must be much smaller than the
frequency difference associated with flip of the second nuclear
spin. The unwanted off-resonant transition is
$|0110\rangle\leftrightarrow|1110\rangle$. The detuning for this
transition is $\delta\omega=2\gamma_n\delta B$. From Eq.~(\ref{mu})
and for $\mu\ll 1$, we obtain the condition
\begin{equation}
\label{nuclearK1}
K_n\gg 363,
\end{equation}
or
$\Omega_2/(2\pi)\ll 162$ kHz.

We now calculate the error associated with displacement of the
first phosphorus atom for $m=-1$.
The corrections to the magnetic field is given
by Eq.~(\ref{dB}). The correction to
$\gamma_n\delta B/(2\pi)$ is 0.86 kHz and the correction to
$\gamma_n b/(2\pi)$ is -0.86 kHz. From Eq.~(\ref{nu2}) the unwanted
detuning from the resonance is
$$
{\Delta^\prime_2\over 2\pi}=-1.72~\text{ kHz}.
$$
From the condition $\Omega_2\gg |\Delta^\prime_2|$, we find
$\Omega_2\gg 1.3$ kHz, or
\begin{equation}
\label{nuclearK2}
K_n\ll 34,148.
\end{equation}
From Eqs.~(\ref{nuclearK1}) and (\ref{nuclearK2}), one can see that
it is possible to make both the errors due to the nonresonant
excitations and the errors due to the unwanted displacements small.

\begin{figure}
%\vspace{-7mm}
\includegraphics[width=10cm,height=8cm]{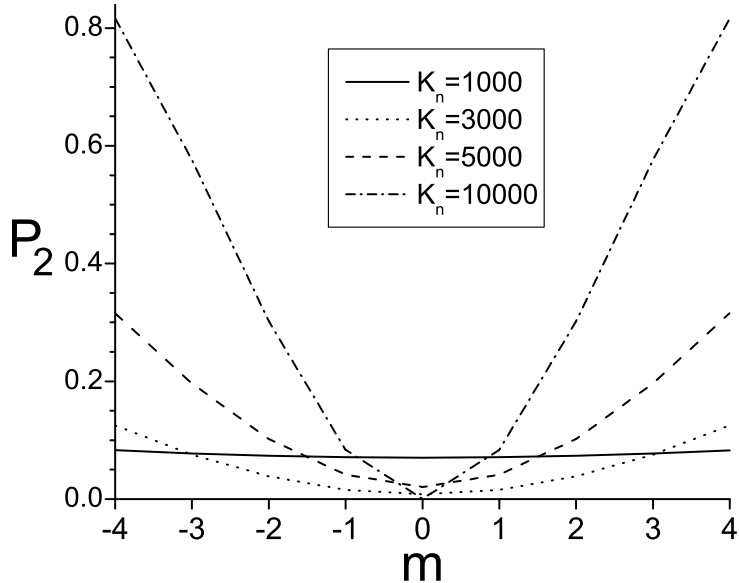}
\vspace{-4mm}
\caption{The probability error $P_2$ generated by the second pulse
on the first nuclear spin as a function of displacement $m$ of the first
phosphorus atom along the $x$ axis for five
values of $K_n$. The data are obtained by numerical
solution of the full Hamiltonian~(\ref{H}).}
\label{reffig:pulse2}
\end{figure}

In Fig.~\ref{reffig:pulse2} we plot the error as a function of
dimensionless displacement $m$ for four different values of $K_n$.
This plot has the same features as Fig.~\ref{reffig:P1}(a).
The probability error $P_2$ is independent of the direction of the
displacement because $P_2$ depends only on the magnitude of 
the deviation of
magnetic field from the optimal value in the location of the qubit.
The error due to the deviation of $J$ from the optimal value $J_0$ does not
contribute to $P_2$ because the first nuclear spin does not directly interact
with the second electron spin and second nuclear spin.
Since the nuclear spins do not interact with each other
and electron and nuclear spins of different phosphorus atoms 
do not interact with each other,
the displacement of the second phosphorus
atom practically does not affect the quantum logic
operations on the nuclear spin of the first atom.
This was also confirmed by our calculations (not presented here).
This result follows also from Eq.~(\ref{nu2}), where 
$\nu_2=\gamma_nB_1^0+A/2$ is independent of $J$ and $B_2^0$.

\begin{figure}
%\vspace{-7mm}
\includegraphics[width=10cm,height=8cm]{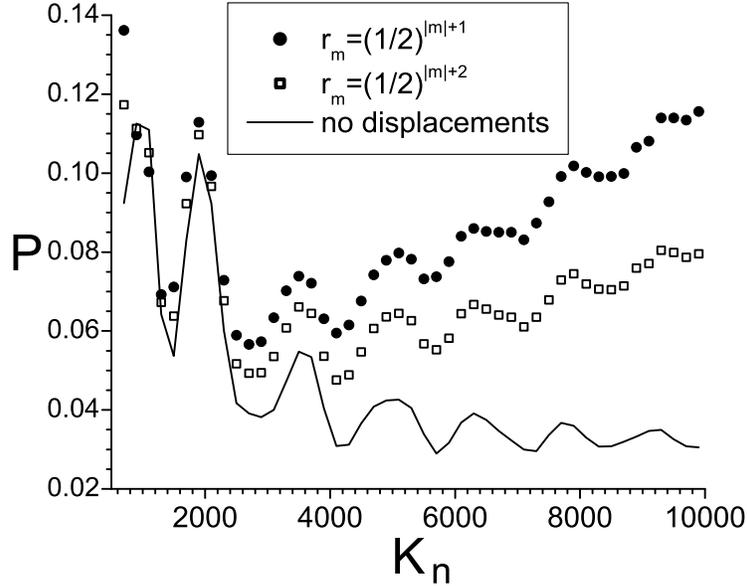}
\vspace{-4mm} \caption{The probability of error $P=1-|C_{15}(T)|^2$
generated by the initialization protocol as a function of $K_n$.
Each phosphorus atom in an ensemble of 10,000 spin pairs had the
probability $r_m$, $m=-4,-3,\dots,3,4$ to be displaced in a random
direction along the $x$ axis. For a perfect sample without
displacements $r_m=0$ (solid line), $K_e=1$. The data are obtained
by numerical solution of the full Hamiltonian~(\ref{H}).}
\label{reffig:init}
\end{figure}

\section{Error in the initialization algorithm}
We have analyzed the errors generated by the first and second 
pulses. The errors generated by the third and fourth pulses [pulses (c) 
and (d) in Eq.~(\ref{Initialization})] are of the same order of magnitude.
As follows from these results, one can make the error relatively small
and polarize (initialize) the nuclear spins
even in the situation when the phosphorus atoms
are displaced from their prescribed locations in the chains.
To illustrate this feature we modeled the initialization protocol
on an ensemble of 10,000 noninteracting spin chains. Each chain
consisted of two phosphorus atoms. The initial state
for each chain was the superposition
$$
C_6(0)|0110\rangle+C_7(0)|0111\rangle
+C_{14}(0)|1110\rangle+C_{15}(0)|1111\rangle
$$
with randomly chosen normalized complex coefficients $C_i(t)$,
$i=6,7,14,15$. The probability of error, $P=1-|C_{15}(T)|^2$, where $T$
is the time of implementation of the initialization algorithm,
as a function of $K_n$ is presented in Fig.~\ref{reffig:init}. We
calculated three situations. In the first situation each phosphorus atom
in the ensemble had the probability $r_m=(1/2)^{|m|+1}$,
$m=-4,-3,\dots,3,4$ to be displaced in a random direction along the
$x$ axis (filled circles in Fig.~\ref{reffig:init}). In the second
situation the probability of displacement is $r_m=(1/2)^{|m|+2}$
which corresponds to a sample of higher quality. For comparison, 
for the third situation we
plot $P$ for a perfect sample without displacements ($r_m=0$). The
data in Fig.~\ref{reffig:init} are averaged over 40 realizations
with different initial conditions and different sets of random
displacements. The size of the error bars is of the order of the
size of the symbols. The Rabi frequencies corresponding to different
values of $K_n$ can be obtained from Eq.~(\ref{Omega2}) with
$|\Delta_2|/(2\pi)\approx|\Delta_4|/(2\pi)\approx 117.5$ MHz. If
$K_n$ assumes the values in the interval from 700 to 10,000, the
magnitudes of $\Omega_2$ and $\Omega_4$ vary in the interval from 84
kHz to 5.9 kHz.

As follows from Fig. 6, when $K_n$ is relatively small (and the Rabi
frequency is relatively large) the error is mostly determined by the
off-resonant excitations and only slightly depends on the quality of the
sample. For relatively large values of $K_n$ (and small Rabi
frequencies) the error is smaller for better samples with smaller
number of displacements.

\section{Controlled NOT gate between electron spins}
As follows from the results presented above, one can  make the error
in the Controlled NOT gates between the nuclear and electron spins of the
same phosphorus to be relatively
small because the coupling constant $A$ responsible for the
Controlled NOT gates is not modified by the displacements.
Now we will consider the error generated by the Controlled NOT gate
between the electron spins
of the neighboring phosphorus atoms. One can expect this error
to be relatively large because a displacement modifies the
coupling constant $J$ which is the key parameter for this gate.

The detuning is $|\Delta_e|\approx J$, so that the Rabi frequency is equal to
\begin{equation}
\label{Omega7}
\Omega_e={|\Delta_e|\over \sqrt{4{K_e^\prime}^2-1}}\approx
{J\over \sqrt{4{K_e^\prime}^2-1}},
\end{equation}
where $K_e^\prime=1,2,\dots$ is an integer number. From Eq.~(\ref{J})
we can find the modification $\delta J=J-J_0$ from the ``ideal'' value $J_0$
due to the displacement $a_0m=7.68m$~\AA,
$$
\delta J\approx J_0\left[{5m\over 2N}+e^{2a_0m/a_{\rm B}}-1\right].
$$
For $N=47$ and $a_B=33.14$~\AA,~we obtain
$$
{\delta J/ J_0}=0.4103m+0.1082m^2+0.0166m^3+0.0019m^4+\dots.
$$
For the minimum possible displacement with $m=-1$, we have
$\delta J/J_0\approx -0.32$ which can be verified from data in Table I.

The unwanted modification of the Rabi frequency (\ref{Omega7})
caused by the displacement is
\begin{equation}
\label{deltaOmega_e}
\delta\Omega_e={\delta J\over \sqrt{4{K_e^\prime}^2-1}}=-0.32\Omega_e.
\end{equation}
The error is small if two conditions are satisfied: (i)
$|\delta\Omega_e|\ll\Omega_e$ and (ii) modulus of the unwanted
detuning $|\Delta^\prime_e|$ is small, $|\Delta^\prime_e|\ll
\Omega_e$. One can show that the unwanted detuning from the
resonance caused by the displacement is
$|\Delta^\prime_e|\approx\delta J/2$, so that both conditions (i)
and (ii) cannot be satisfied by any choice of parameters of the
system and the error is large. We checked this result numerically
and determined that the error $P_e$ is close to unity for $|m|\ne 0$
and $|P_e|\sim 10^{-4}$ to $10^{-3}$ for $m=0$. 

We now estimate the error
if one qubit is displaced in a spin chain with more than two phosphorus atoms. 
The presented above analysis is valid for the edge qubits of this chain.
Consider, for example, the state 
$|\cdots 1_{k+1}1_{k}1_{k-1}\cdots\rangle$
formed by coupled electron spins. (Here we do not indicate the states 
of the nuclear spins.) If the the $k$th atom is not displaced, then the 
frequency resonant for the transition of the $k$th qubit is 
\begin{equation}
\label{nue}
{\nu_e\over 2\pi}={1\over 2\pi}\left({J_0\over 2}+{J_0\over 2}+\dots\right)
={J_0\over 2\pi}+\dots, 
\end{equation}
where $J_0/(2\pi)=1.97$ MHz.
If the $k$th qubit is displaced 
in any direction along the chain by one lattice cite then instead of 
the term $J_0/(2\pi)$
in the right-hand side of Eq.~(\ref{nue}) we have from Table I
$$
{1\over 2\pi}\left[{J(m=1)\over 2}+{J(m=-1)\over 2}\right]=2.14~\rm{MHz},
$$
so that the unwanted detuning from the resonance is 
\begin{equation}
\label{CNee1}
\Delta_e^\prime=0.086J_0.
\end{equation}
From Eq.~(\ref{Omega7}) for $K_e^\prime=1$, we obtain 
\begin{equation}
\label{CNee2}
\Omega_e={J_0\over \sqrt{3}}\approx 0.58J_0  
\end{equation}
From Eqs.~(\ref{CNee1}) and (\ref{CNee2}) one can see that the 
unwanted detuning is relatively small $|\Delta_e^\prime|\ll \Omega_e$
and condition (ii) is satisfied. One can show that this condition is 
not satisfied for the transitions 
$|\cdots 1_{k+1}1_{k}0_{k-1}\cdots\rangle
\leftrightarrow|1_{k+1}0_{k}0_{k-1}\rangle$.
The deviation in Rabi frequency [condition (i)] is always relatively large. 
For example, if the target $k$th qubit is displaced in the negative direction 
(m=-1) and the $(k-1)$th qubit is the control qubit,
then $\delta\Omega_e$ is given by Eq.~(\ref{deltaOmega_e})
and condition $|\delta\Omega_e|\ll\Omega_e$ is not satisfied, 
so that the Controlled NOT gate between the displaced electron
spins generate large error. 

Presented here analysis demonstrate that the Controlled NOT gate 
between the electron spins
cannot be performed on displaced qubits. If quantum
computation is implemented on an ensemble of many spin chains, the
errors can be small if the number of chains with displaced qubits is small.

In conclusion, we note that the problem of unwanted displacements appears
only in quantum computation performed
on an ensemble of spin chains. If one works
with one chain only, one can measure the positions of the phosphorus atoms
using a scanning tunneling microscope and optimize the parameters of the
pulses appropriately.

\section*{Acknowledgments}
We thank M. E. Hawley, G. W. Brown, and G. D. Doolen for useful 
discussions.
This work was carried out under the auspices of the National Nuclear
Security Administration of the U.S. Department of Energy at Los
Alamos National Laboratory under Contract No. DE-AC52-06NA25396.

\appendix
\section*{Appendix}
The eigenvalues of the system of
two phosphorus atoms in a permanent magnetic field are
$$
E_0=(\gamma_e-\gamma_n)b+\frac A2+\frac J4,
$$
$$
E_1^{(0)}=\gamma_eb-\gamma_n\delta B+\frac J4,~~~
E_1^{(2)}={A^2\over 4\left[E_1^{(0)}-E_2^{(0)}\right]},
$$
$$
E_2^{(0)}=-\gamma_nb-\frac J4+\sqrt{(\gamma_e\delta B)^2+{J^2\over 4}},~~~
E_2^{(2)}=-E_1^{(2)},
$$
$$
E_3^{(0)}=-\gamma_n\delta B-\frac J4+
\sqrt{\left(\gamma_e\delta B+\frac A2\right)^2+{J^2\over 4}},~~~
E_3^{(2)}=0,
$$
$$
E_4^{(0)}=-\gamma_nb-\frac J4-\sqrt{(\gamma_e\delta B)^2+{J^2\over 4}},~~~
E_4^{(2)}={A^2\over 4\left[E_4^{(0)}-E_8^{(0)}\right]},
$$
$$
E_5^{(0)}=-\gamma_n\delta B-\frac J4-
\sqrt{\left(\gamma_e\delta B+\frac A2\right)^2+{J^2\over 4}},~~~
$$
$$
E_5^{(2)}={A^2\over 4}\left[
{1\over E_5^{(0)}-E_{6}^{(0)}}+{1\over E_5^{(0)}-E_{9}^{(0)}}\right],
$$
$$
E_6^{(0)}=-(\gamma_e+\gamma_n)b-\frac A2+\frac J4,~~~
E_6^{(2)}={A^2\over 4}\left[
{1\over E_6^{(0)}-E_{5}^{(0)}}+{1\over E_6^{(0)}-E_{10}^{(0)}}\right],
$$
$$
E_7^{(0)}=-\gamma_eb-\gamma_n\delta B+\frac J4,~~~
E_7^{(2)}={A^2\over 4\left[E_7^{(0)}-E_{11}^{(0)}\right]},
$$
$$
E_8^{(0)}=\gamma_eb+\gamma_n\delta B+\frac J4,~~~
E_8^{(2)}=-E_4^{(2)},
$$
$$
E_9^{(0)}=(\gamma_e+\gamma_n)b-\frac A2+\frac J4,~~~
E_9^{(2)}={A^2\over 4}\left[
{1\over E_9^{(0)}-E_{5}^{(0)}}+{1\over E_9^{(0)}-E_{10}^{(0)}}\right],
$$
$$
E_{10}^{(0)}=\gamma_n\delta B-\frac J4+
\sqrt{\left(\gamma_e\delta B-\frac A2\right)^2+{J^2\over 4}},~~~
$$
$$
E_{10}^{(2)}={A^2\over 4}\left[
{1\over E_{10}^{(0)}-E_{6}^{(0)}}+{1\over E_{10}^{(0)}-E_{9}^{(0)}}\right],
$$
$$
E_{11}^{(0)}=\gamma_nb-\frac J4+\sqrt{(\gamma_e\delta B)^2+{J^2\over 4}},~~~
E_{11}^{(2)}=-E_{7}^{(2)},
$$
$$
E_{12}^{(0)}=\gamma_n\delta B-\frac J4-
\sqrt{\left(\gamma_e\delta B-\frac A2\right)^2+{J^2\over 4}},~~~
E_{12}^{(2)}=0,
$$
$$
E_{13}^{(0)}=\gamma_nb-\frac J4-\sqrt{(\gamma_e\delta B)^2+{J^2\over 4}},~~~
E_{13}^{(2)}={A^2\over 4\left[E_{13}^{(0)}-E_{14}^{(0)}\right]},
$$
$$
E_{14}^{(0)}=-\gamma_eb+\gamma_n\delta B+\frac J4,~~~
E_{14}^{(2)}=-E_{13}^{(2)},
$$
$$
E_{15}=(-\gamma_e+\gamma_n)b+\frac A2+\frac J4,
$$
where $b=(B_2^0+B_1^0)/2$ and $\delta B=(B_2^0-B_1^0)/2$.
In the text we assume $E_i=E^{(0)}_i+E^{(2)}_i$.
The eigenvalues $E_{10}$ and $E_{12}$
are presented for the case $A/2\le \gamma_e\delta B$. For the opposite case,
$A/2>\gamma_e\delta B$, one must exchange the eigenvalues
$E_{10}\leftrightarrow E_{12}$.


\begin{thebibliography}{}
\bibitem{1994} J. W. Lyding, G. C. Abeln, T.-C. Shen, C. Wang,
and J. R. Tucker,
%Nanometer scale patterning and oxidation in silicon surfaces with
%an ultrahigh vacuum scanning tunneling microscope,
J. Vac. Sci. Technol. B {\bf 12}, 3735 (1994).
\bibitem{1996} D. P. Adams, T. M. Mayer, and B. S. Swartzentruber,
%Nanometer-scale lithography on Si(001) using adsorbed H as an
%atomic layer resist,
J. Vac. Sci. Technol. B {\bf 14}, 1642 (1996).
\bibitem{surfSci}
C. Thirstrup, M Sakurai, T. Nakayama, M. Aono,
%Atomic scale modification of hydrogen-terminated silicon 2$\times$1
%and 3$\time$1 (001) surfaces by scanning tunneling microscope,
Surf. Sci. {\bf 411}, 203 (1998).
\bibitem{electronics} J. R. Tucker, and T.-C. Shen,
% Prospects for atomically ordered device structures based on STM
% lithography,
Solid-State Electronics {\bf 42}, 1061 (1998).
\bibitem{clark} J. L. O'Brien, S. R. Schofield, M. Y. Simmons, R. G. Clark,
A. S. Dzurak, N. J. Curson, B. E. Kane, N. S. McAlpine, M. E. Hawley,
and G. W. Brown,
%Towards the fabrication of phosphorus qubits for a silicon quantum computer,
Phys. Rev. B {\bf 64}, 161401(R) (2001).
\bibitem{Kane} B. E. Kane, Nature (London) {\bf 393}, 133 (1998)
\bibitem{optical1} J. Kohler,
%Magnetic resonance of a single molecular spin,
Phys. Rep. {\bf 310}, 261 (1999).
\bibitem{optical2} S. Ya. Kilin, A. P. Nizovtsev, T. M. Maevskaya,
A. Drabenstedt, and J. Wrachtrup,
%Spectroscopy of single N-V defect centers in diamond: tunneling
%of nitrogen atoms into vacancies and fluorescence spectra,
J. Luminescence {\bf 86}, 201 (2000).
\bibitem{optical3} F. T. Charnock and T. A. Kennedy,
% Combined optical and microwave approach for performing quantum spin
%operations on the nitrogen-vacancy center in diamond,
Phys. Rev. B {\bf 64}, 041201(R) (2001).
\bibitem{T_c} A. M. Tyryshkin, S. A. Lyon, A. V. Astashkin,
and A. M. Raitsimring,
%Electron spin relaxation times of phosphorus donors in silicon
Phys. Rev. B {\bf 68}, 193207 (2003).
\bibitem{obus1} J. I. Cirac, P. Zoller, H. J. Kimble, and H. Mabuchi, 
%Quantum state transfer and entanglement distribution among 
%distant nodes in a quantum network,
Phys. Rev. Lett. {\bf 78}, 3221 (1997).
\bibitem{obus2} D. E. Browne, M. B. Plenio, S. F. Huelga, 
%Robust creation of entanglement between ions in spatially separate cavities
Phys. Rev. Lett. {\bf 91}, 067901 (2003).
\bibitem{obus3} S. Clark, A. Peng, M. Gu, and S. Parkins, 
%Unconditional Preparation of Entanglement between Atoms in 
% Cascaded Optical Cavities
Phys. Rev. Lett. {\bf 91}, 177901 (2003).
\bibitem{obus4} S. Mancini and S. Bose,  
%Engineering an interaction and entanglement between distant atoms,
Phys. Rev. A {\bf 70}, 022307 (2004).
\bibitem{obus5} L.-M. Duan, B. B. Blinov, D. L. Moehring, and C. Monroe, 
%Scalable trapped ion quantum computation with a probabilistic ion-photon
%mapping, 
Quant. Inf. Comput. {\bf 4}, 165 (2004).
\bibitem{obus6} Y. L. Lim, A. Beige, and L. C. Kwek, 
% Repeat-until-success linear optics distributed quantum computing,
Phys. Rev. Lett. {\bf 95}, 030505 (2005).
\bibitem{obus7} L.-M. Duan, B. Wang, H. J. Kimble, 
%Robust quantum gates on neutral atoms with cavity-assisted 
%photon scattering,
Phys. Rev. A {\bf 72}, 032333 (2005).
\bibitem{xxx}
G. P. Berman, G. W. Brown, M. E. Hawley, D. I. Kamenev, and V. I. Tsifrinovich,
%Implementation of quantum logic operations and creation of
%entanglement in a silicon-based quantum computer with constant interaction,
quant-ph/0512174.
\bibitem{feher} G. Feher,
% Electron spin resonance experiments on donors in silicon.
% I. Elecronic structure of donors by the electron nuclear double
%resonance technique,
Phys. Rev. {\bf 114}, 1219 (1959).
\bibitem{book}
G. P. Berman, G. D. Doolen, R. Mainieri, and V. I. Tsifrinovich,
{\it Introduction to Quantum Computers} (World Scientific, Singapore, 1998).
\bibitem{book1}
G. P. Berman, D. I. Kamenev, and V. I. Tsifrinovich,
{\it Perturbation Theory for Solid-State Quantum Computation with Many
Quantum Bits} (Rinton Press, Princeton, 2005).
\bibitem{HerringFlicker} C. Herring, M. Flicker,
%Asymptotic exchange coupling of two hydrogen atoms,
Phys. Rev. {\bf 134}, A362 (1964).
\bibitem{PR02} G. P. Berman, G. D. Doolen, D. I. Kamenev, and
V. I. Tsifrinovich,
%Perturbation theory for quantum computation with large number of qubits,
Phys. Rev. A {\bf 65}, 012321  (2002).
\end{thebibliography}
\end{document}